\documentstyle[aps,prb,floats]{revtex}

\begin{document}
\input epsf
\draft
\renewcommand{\floatpagefraction}{1.00}
\renewcommand{\topfraction}{1.00}
\renewcommand{\textfraction}{0.00}
\renewcommand{\bottomfraction}{1.00}
\twocolumn[\hsize\textwidth\columnwidth\hsize\csname@twocolumnfalse%
\endcsname
\title{  {\rm\small\hfill (submitted to Phys. Rev. B})\\
Composition, structure and stability of RuO${}_2$(110)\\
as a function of oxygen pressure}

\author{Karsten Reuter and Matthias Scheffler}

\address{Fritz-Haber-Institut der Max-Planck-Gesellschaft, Faradayweg
4-6, D-14195 Berlin-Dahlem, Germany}

\date{Received 11 July 2001}

\maketitle

\begin{abstract}
Using density-functional theory (DFT) we calculate the Gibbs free
energy to determine the lowest-energy structure of a RuO${}_2$(110)
surface in thermodynamic equilibrium with an oxygen-rich environment.
The traditionally assumed stoichiometric termination is only found
to be favorable at low oxygen chemical potentials, i.e. low pressures
and/or high temperatures. At realistic O pressure, the surface is
predicted to contain additional terminal O atoms. Although this O
excess defines a so-called polar surface, we show that the prevalent
ionic model, that dismisses such terminations on electrostatic grounds,
is of little validity for RuO${}_2$(110). Together with analogous results
obtained previously at the (0001) surface of corundum-structured oxides,
these findings on (110) rutile indicate that the stability of
non-stoichiometric terminations is a more general phenomenon on
transition metal oxide surfaces.
\end{abstract}

\hfill {\quad}
]

\section{Introduction} 
Density-functional theory (DFT) is often argued to be a zero temperature,
zero pressure technique. As such, the results of static total energy
calculations at surfaces have to be transferred with considerable care to
typical high-pressure applications like e.g. catalysis --- a situation which 
finds its correspondence in the experiment in form of the {\em ex-situ}
methods of ultra high vacuum (UHV) surface science. Unfortunately, often
enough an extrapolation of the low-pressure results to technical processes
taking place at ambient atmosphere is not possible at all, which has been
coined with buzz words like {\em pressure} and {\em materials gap}
(e.g. see the discussion in Stampfl {\em et al.} \cite{stampfl01} and
references therein).

Trying to bridge these gaps, one needs to determine the equilibrium composition
and geometry of a surface in contact with a given environment at finite
temperature and pressure. Then, the stable surface structure results from
the statistical average over adsorption and desorption processes, and this
means to employ thermodynamics. When we aim to describe experiments that are
conducted at constant pressure and temperature, the appropriate thermodynamic
potential to consider is the Gibbs free energy, $G(T,p)$. If DFT total
energies enter in a suitable way into the calculation of $G(T,p)$ for a
material surface, an {\em ab initio thermodynamics} results, which extends
the predictive power of the first-principles technique to a more relevant
temperature and pressure range.

This scheme has been successfully applied to e.g. address
the surface termination of corundum-structured oxides \cite{wang98,wang00},
and we will use it here to determine the composition and lowest energy
structure of RuO${}_2$(110) in equilibrium with an oxygen atmosphere.
At oxide surfaces typically only so-called autocompensated, stoichiometric
terminations are considered \cite{henrich94,noguera96}, i.e. they are
believed to be more favorable than the other, so-called polar terminations
\cite{noguera00} for two reasons: First, they often involve a minimum
of truncated bonds at the surface, and, second, in a purely electrostatic
model, in which all oxide ions would be in their bulk formal oxidation
state, polar surfaces would be charged and should thus exhibit an infinite
surface energy.

In contrast to these arguments, our {\em ab initio thermodynamics}
calculations show that on RuO${}_2$(110) a polar surface termination
with excess oxygens is stabilized at higher O chemical potentials.
Hence, depending on the experimental conditions, either the stoichiometric
UHV or the hitherto unaccounted high-pressure termination can be present,
which must be considered in the modelling of physical processes
occuring at this surface, as e.g. catalytic reactions. We will also show
that the rejection of polar surfaces on electrostatic grounds is not valid,
as the strong dipole moment can be considerably reduced by surface
relaxation and electron rearrangement: Rather than imagining the surface as
created by a truncation of the bulk-stacking sequence at a certain bulk-like
plane (which is the basis of the electrostatic divergence argument), one
should instead view the surface as a new material, in which the structural
and electronic degrees of freedom of the top atomic layers allow a significant
modification of the properties of the atoms in the bulk.

These results for the rutile-structured RuO${}_2$(110) are analogous to
previous findings at the (0001) surface of corundum-structured oxides 
\cite{wang98,wang00}, supporting the argument that polar terminations,
particularly at realistic pressure, are a more general phenomenon on
transition metal oxide surfaces.

\section{Theoretical}

This Section describes the thermodynamic formalism and how it is combined
with DFT total energy calculations. For the sake of clarity this is all
referred explicitly to the present application to RuO${}_2$(110) in an
oxygen atmosphere. However, the generalization to other compounds,
M${}_{\rm x}$O${}_{\rm y}$, and even to an environment, that contains
multiple gas phase species and not just oxygen, is obvious.

\subsection{The surface free energy} 

We consider a surface in contact with an oxygen atmosphere described
by an oxygen pressure, $p$, and temperature, $T$. This means that the
environment acts as a reservoir, because it can give (or take) any
amount of oxygen to (or from) the sample without changing the temperature
or pressure. The appropriate thermodynamic potential required to
describe such a system is the Gibbs free energy, $G(T,p,N_{\rm Ru},
N_{\rm O})$, which depends also on the number of Ru, $N_{\rm Ru}$, and
O, $N_{\rm O}$, atoms in the sample. The most stable surface composition
and geometry is then the one that minimizes the surface free energy,
$\gamma(T,p)$, defined as

\begin{eqnarray}
\label{gammafund}
\gamma(T,p) \;=\; \frac{1}{A} && \left[ \;\; G(T,p,N_{\rm Ru},N_{\rm O})
\;-\; \right. \\ \nonumber
                               &-& \left. N_{\rm Ru} \mu_{\rm Ru}(T,p) \;-\; N_{\rm O}  \mu_{\rm O}(T,p) \;\; \right].
\end{eqnarray}

\noindent
Here, $\mu_{\rm Ru}$ and $\mu_{\rm O}$ are the chemical potentials of a Ru and
an O atom, respectively, and $\gamma(T,p)$ is normalized to energy per unit
area by dividing through the surface area, $A$.

If the surface system is modeled by a slab with two equivalent surfaces,
eq. (\ref{gammafund}) reads

\begin{eqnarray}
\label{gammastart}
\gamma(T,p) \;=\; \frac{1}{2A} && \left[ \;\; G^{\rm slab}(T,p,N_{\rm Ru},N_{\rm O})
\;-\; \right. \\ \nonumber
                               &-& \left. N_{\rm Ru} \mu_{\rm Ru}(T,p) \;-\; N_{\rm O}  \mu_{\rm O}(T,p) \;\; \right].
\end{eqnarray}

\noindent
Now, $A$ is the area of the surface unit cell and $N_{\rm Ru}$ and
$N_{\rm O}$ are the numbers of Ru and O atoms in the three-dimensional
super cell.

In eq. (\ref{gammastart}) the chemical potentials of O and
Ru enter in a symmetric way. However, if there is enough bulk material,
so that it acts as a thermodynamic reservoir, the potentials are in fact
no longer independent, but are related by the Gibbs free energy of the
bulk oxide

\begin{equation}
\mu_{\rm Ru}(T,p) \;+\; 2 \mu_{\rm O}(T,p) \;=\; g^{\rm bulk}_{\rm RuO_2}(T,p),
\label{bulkequilibrium}
\end{equation} 

\noindent
where lower case $g$ is henceforth used to denote a Gibbs free energy
per formula unit. Inserting this constraint into eq. (\ref{gammastart})
leads to

\begin{eqnarray}
\label{gammabasic}
\gamma(T,p) &=& \frac{1}{2A} \bigg[ \; G^{\rm slab}(T,p,N_{\rm Ru},N_{\rm O})
\;-\; \\ \nonumber
            && N_{\rm Ru}\; g^{\rm bulk}_{\rm RuO_2}(T,p) \;+\;
               \left( 2 N_{\rm Ru} - N_{\rm O} \right) \mu_{\rm O}(T,p) \bigg] ,
\end{eqnarray}

\noindent
which shows how the surface free energy depends now only on the oxygen
chemical potential.

\subsection{The range of allowed O chemical potentials}

As suggested by experimental conditions, we like to use the oxygen
chemical potential to discuss the dependence of the surface on the
O${}_2$ pressure and temperature. It is important to note, that
experimentally (and assuming that thermodynamic equilibrium applies)
$\mu_{\rm O}$ cannot be varied without bounds. If it gets too low,
all oxygen would leave the sample, i.e. the oxide would decompose
into solid Ru and oxygen gas, which would start with the formation
of Ru crystallites at the surface. Thus,

\begin{equation}
max\big(\mu_{\rm Ru}(T,p) \big) \;=\; g^{\rm bulk}_{\rm Ru}(T,p),
\end{equation}

\noindent
where $g^{\rm bulk}_{\rm Ru}(T,p)$ is the Gibbs free energy of
metallic ruthenium. Together with eq. (\ref{bulkequilibrium})
and using the $T= 0$\,K and $p = 0$\,atm limit for the bulk energies,
we will employ

\begin{equation}
min\big(\mu_{\rm O}(T,p) \big) \;\stackrel{!}{=}\;
\frac{1}{2} \left( \; g^{\rm bulk}_{\rm RuO_2}(0,0)
\;-\; g^{\rm bulk}_{\rm Ru}(0,0) \; \right),
\end{equation}

\noindent
to mark the "oxygen-poor limit" (or equivalently "Ru-rich" limit)
in the graphs discussed below. This is a good estimate of the
real limit and most importantly, it is a theoretically well defined
reference-point on the $\mu_{\rm O}$-axis.

On the other hand, the most oxygen-rich conditions can be defined
as the point, beyond which gas phase O would start to condensate on
the sample. However, in the temperature and pressure range we are
interested in a condensed O${}_2$-solid phase does not exist (the
critical temperature of O${}_2$, i.e. above which gas and liquid
phase are degenerate, is $T_c \approx 150$ K). Thus, similarly as
above, an appropriate and well-defined estimate of the upper limit
of the oxygen chemical potential is

\begin{equation}
max\big(\mu_{\rm O}(T,p) \big) \;\stackrel{!}{=}\;
1/2 \;E^{\rm total}_{{\rm O}_2},
\end{equation}

\noindent
where $E^{\rm total}_{\rm O_2}$ is the total energy of a free, isolated
O${}_2$ molecule at $T = 0$ K.

Then, introducing the Gibbs free energy of formation, $\Delta G_f(T,p)$,
of the oxide,

\begin{equation}
\Delta G_f(T,p) \;=\; g^{\rm bulk}_{\rm RuO_2}(T,p) - g^{\rm bulk}_{\rm Ru}(T,p) - g^{\rm gas}_{\rm O_2}(T,p) ,
\end{equation}

\noindent
where $g^{\rm gas}_{\rm O_2}(T,p)$ is the Gibbs free energy of an
O${}_2$ molecule, we see that the range of oxygen chemical potentials
between our theoretical tick marks is

\begin{equation}
\frac{1}{2} \Delta G_f(0,0) \;<\;
\mu_{\rm O}(T,p) - \frac{1}{2} \; E^{\rm total}_{\rm O_2} \;<\; 0.
\label{oxygenlimits}
\end{equation}

\noindent
We compute $\Delta G_f(0,0) = - 3.16$ eV per formula unit, which
compares very well with the experimental Gibbs free energy of formation
at standard pressure in the limit of low temperatures,
$\Delta G_f^o(T \rightarrow 0{\rm K}, 1{\rm atm}) = - 3.19$ eV
per formula unit \cite{CRC95}.

It is important to notice, that our "tick marks" for the oxygen-rich
and oxygen-poor conditions are theoretically well defined limits, yet
they only represent an estimate of the truly accessible range of
the oxygen chemical potential. The range between our tick marks is
$1/2\; \Delta G_f(0,0)$, but in reality the accessible range
is $1/2\; \Delta G_f(T,p)$, i.e. it is temperature and pressure
dependent. At $T = 1000$\,K and $p = 1$\,atm the Gibbs free energy of
formation has increased by 0.63\,eV compared to the aforementioned
$T \rightarrow 0$\,K value \cite{CRC95}. Keeping the ensuing
variation of 0.3\,eV in the accessible range of potentials in mind,
we will always show the resulting curves in our below discussed
figures also some tenths of an eV outside the "oxygen-rich" and
"oxygen-poor" tick marks.

\subsection{The oxygen-poor limit as a safe reference}

Total energies for extended systems are typically more accurately
described by actual DFT calculations than those for atoms and
molecules. With respect to eq. (\ref{oxygenlimits}), it is therefore
suitable to rewrite it to

\begin{eqnarray}
\label{oxygenlimits3}
\frac{1}{2} \big( g^{\rm bulk}_{\rm RuO_2}(0,0) - g^{\rm bulk}_{\rm Ru}(0,0) \big)
&<& \mu_{\rm O}(T,p) \;<\; \\ \nonumber
\frac{1}{2} \big( g^{\rm bulk}_{\rm RuO_2}(0,0) - g^{\rm bulk}_{\rm Ru}(0,0) \big)
&+& \frac{1}{2} \Delta G_f(0,0).
\end{eqnarray}

\noindent
If we then insert the oxygen-poor limit into eq. (\ref{gammabasic}) we
obtain for the surface free energy

\begin{eqnarray}
\label{gammamrich}
\gamma_{\rm O-poor}&&(T,p) \;= \\ \nonumber
\frac{1}{2A} && \bigg[ \; G^{\rm slab}(T,p,N_{\rm Ru},N_{\rm O})
\;-\; N_{\rm Ru} g^{\rm bulk}_{\rm RuO_2}(T,p) \;-\;  \\ \nonumber
&& \left( N_{\rm Ru} - \frac{N_{\rm O}}{2} \right)
\big( g^{\rm bulk}_{\rm Ru}(0,0) \;-\; g^{\rm bulk}_{\rm RuO_2}(0,0)
\big) \; \bigg].
\end{eqnarray}

\noindent
Likewise, the oxygen-rich limit turns out to be

\begin{eqnarray}
\label{gammaorich}
\gamma_{\rm O-rich}(T,p) &=& \gamma_{\rm O-poor}(T,p) \;-\; \\ \nonumber
                         &&  \frac{1}{2A} \left( N_{\rm Ru} - \frac{N_{\rm
                         O}}{2} \right) \Delta G_f(0,0). 
\end{eqnarray}

\noindent
The result of this rewriting of eq. (\ref{oxygenlimits}) is that
atomic or molecular quantities do not enter into the calculation of
the oxygen-poor limit, i.e. eq. (\ref{gammamrich}), at all, which thus
defines a safe reference involving only bulk or slab quantities.

On the other hand, $\Delta G_f(0,0)$ depends on the O${}_2$ total
energy, and $\Delta G_f(0,0)$ defines the slope of the lines representing
the surface free energy as a function of $\mu_{\rm O}$: The slope
is

\begin{equation}
\frac{1}{2A} \left( N_{\rm Ru} - \frac{N_{\rm O}}{2} \right) \Delta G_f(0,0),
\end{equation}

\noindent
cf. eq. (\ref{gammaorich}), and sometimes $\Delta G_f(0,0)$ may be
affected by the error in $E^{\rm total}_{\rm O_2}$, in which case
it might be preferable to use its experimental value. Yet, for the
present case of RuO${}_2$, we note that our DFT-GGA result for the
Gibbs free energy of formation is very close to the experimental
value (s. above). Thus, here the error in $E^{\rm total}_{\rm O_2}$,
which clearly exists, cancels out and the calculated slopes are
very accurate. As a consequence, and in contrast to common belief,
we note that the bulk total energy of RuO${}_2$ must therefore have
a similar error as $E^{\rm total}_{\rm O_2}$ -- otherwise the
apparent error cancelation in $\Delta G_f(0,0)$ would not occur.

We finally note in passing, that eqs. (\ref{gammamrich}) and
(\ref{gammaorich}) nicely reflect the physics behind the
dependence on the O chemical potential: While a stoichiometrically
terminated surface structure ($N_{\rm Ru} = N_{\rm O}/2$) will
exhibit a constant surface free energy as a function of 
$\mu_{\rm O}(T,p)$, a termination with an O excess (deficiency) will
become more and more favorable (unfavorable) with increasing
$\mu_{\rm O}(T,p)$, i.e. higher O pressure and/or lower temperature.

\subsection{Gibbs free energies vs. total energies}

The formalism as described up to this point is entirely
based on the Gibbs free energies of the system, whereas we intend
to provide as input total energies from DFT calculations. Therefore,
we will now outline how both quantities are related, and
under which approximations they might be equated to each other.

DFT total energies are evaluated for a certain volume, $V$, of
the unit cell. The resulting $E^{\rm total}(V,N_{\rm Ru},N_{\rm O})$
is related to a thermodynamical quantity only in a restricted way:
It corresponds to the Helmholtz free energy at zero temperature and
neglecting zero-point vibrations. In general, the Helmholtz free
energy can thus be written as

\begin{eqnarray}
\label{helmholtz}
F(T,V,N_{\rm Ru},N_{\rm O}) &=& E^{\rm total}(V,N_{\rm Ru},N_{\rm O}) \;+ \\ \nonumber
                            & & F^{\rm vib.}(T,V,N_{\rm Ru},N_{\rm O}) ,
\end{eqnarray}

\noindent
with

\begin{eqnarray}
\label{ffreq}
F^{\rm vib.}(T,V,N_{\rm Ru},N_{\rm O}) &=& E^{\rm vib.}(T,V,N_{\rm Ru},N_{\rm O}) \;- \\ \nonumber
                                       & & T S^{\rm vib.}(T,V,N_{\rm Ru},N_{\rm O})
\end{eqnarray}

\noindent
comprising all contributions, which depend on vibrational modes in
the system. Here, $E^{\rm vib.}$ and $S^{\rm vib.}$ are the vibrational
energy (including the zero-point energy) and entropy respectively. In turn,
the Helmholtz free energy is associated to the Gibbs free energy
via

\begin{eqnarray}
\label{gibbs}
G(T,p,N_{\rm Ru},N_{\rm O}) &=& F(T,p,N_{\rm Ru},N_{\rm O}) \;+\\ \nonumber
                            & & p V(T,p,N_{\rm Ru}, N_{\rm O}).
\end{eqnarray}

Checking first on the $pV$ term, we find from a simple dimensional
analysis, that its contribution to the surface free energy (normalized
to the surface area) will be [$pV / A$] = atm {\AA}${}^3$ / {\AA}${}^2$
$\sim 10^{-8}$ meV/{\AA}${}^2$. As we are only interested in a pressure
range that will not exceed about 100 atm, this contribution is completely
negligible compared to the Helmholtz free energy, which is of the order
of meV/{\AA}${}^2$.

This leaves as the only additional contribution to
$G(T,p,N_{\rm Ru},N_{\rm O})$ apart from the DFT total energy only
the vibrational term $F^{\rm vib.}(T,V,N_{\rm Ru},N_{\rm O})$.
Using the phonon density of state (DOS), $\sigma(\omega)$, this
vibrational part of the free energy can be written as an integral
over the modes, $\omega$,

\begin{equation}
\label{sumfreq}
F^{\rm vib.}(T,V,N_{\rm Ru},N_{\rm O}) \;=\; 
\int d\omega \; F^{\rm vib.}(T,\omega) \; \sigma(\omega),
\end{equation}

\noindent
where an analytic expression for $F^{\rm vib.}(T,\omega)$ is given
in the appendix.

Plugging this into eq. (\ref{gammamrich}) we obtain for the vibrational
contribution to the surface free energy of a stoichiometric termination
($N_{\rm Ru} = N_{\rm O}/2$) at the O-poor limit,

\begin{eqnarray}
\label{gammamvib}
&&\gamma_{\rm O-poor}^{\rm vib.}(T,V) \;=\; \\ \nonumber
&&\frac{1}{2A} \; \int d\omega \; F^{\rm vib.}(T,\omega) \;
\big( \sigma^{\rm slab}(\omega) - N_{\rm Ru} \sigma^{\rm bulk}_{\rm RuO_2}(\omega) \big).
\end{eqnarray}

\noindent
To get an estimate of its value, we use the Einstein model and
approximate the phonon DOS by just one characteristic frequency
for each atom type. If we further consider, that the vibrational
mode of the topmost layer of Ru and of O might be significantly
changed at the surface, we thus have $\bar{\omega}_{\rm O}^{\rm bulk}$
and $\bar{\omega}_{\rm Ru}^{\rm bulk}$ as characteristic frequencies
of O and Ru in RuO${}_2$ bulk, as well as $\bar{\omega}_{\rm O}^{\rm surf.}$
and $\bar{\omega}_{\rm Ru}^{\rm surf.}$ as the respective modes at
the surface. With this simplified phonon DOS, eq. (\ref{gammamvib})
reduces to

\begin{eqnarray}
\label{freqopoor}
\gamma_{\rm O-poor}^{\rm vib.}(T,V) &\approx& \\ \nonumber
\frac{3}{2A}
  \bigg[ & &\left( F^{\rm vib.}(T,\bar{\omega}_{\rm Ru}^{\rm surf.})
- F^{\rm vib.}(T,\bar{\omega}_{\rm Ru}^{\rm bulk}) \right) \;+\; \\ \nonumber
& &         \left( F^{\rm vib.}(T,\bar{\omega}_{\rm O}^{\rm surf.})
- F^{\rm vib.}(T,\bar{\omega}_{\rm O}^{\rm bulk}) \right) \;\bigg].  \nonumber
\end{eqnarray}

\begin{figure}
       \epsfxsize=0.49\textwidth \centerline{\epsfbox{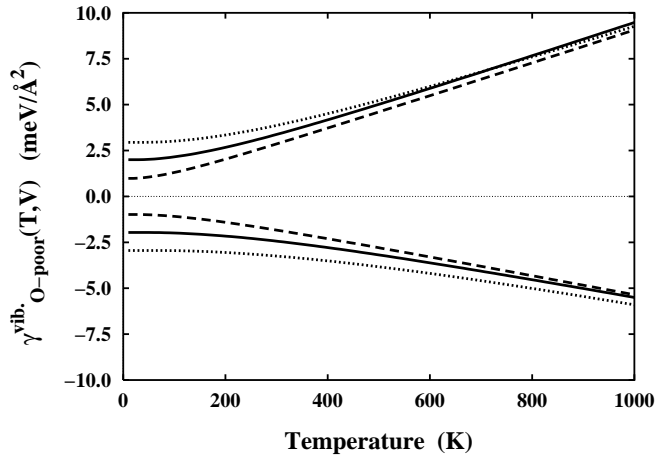}}
       \caption{Vibrational contribution to the surface free energy,
        of a stoichiometric termination, cf. eq. (\ref{freqopoor}), 
        in the temperature range of interest to the present study.
        The Ru and O modes are approximated in the Einstein model
        by characteristic frequencies, $\bar{\omega}_{\rm O}^{\rm bulk}
        = 80$ meV and $\bar{\omega}_{\rm Ru}^{\rm bulk} = $25 meV.
        Shown is the contribution, if the vibrational modes at the
        surface differed by $\pm 50$\% from these bulk values (solid
        lines). To assess the dependence on the value chosen for the
        characteristic frequencies, the latter are varied by $\pm 50$\%
        (dashed and dotted lines respectively). In all cases, the
        vibrational contribution stays below 10 meV/{\AA}${}^2$ in
        the whole temperature range considered. \label{frequency}}
\end{figure}

Hence, we see that in the O-poor limit $\gamma_{\rm O-poor}^{\rm vib.}(T,p)$
of a stoichiometric termination arises primarily out of the difference
of the vibrational modes at the oxide surface with respect to their
bulk value. To arrive at numbers, we use $\bar{\omega}_{\rm O}^{\rm bulk}
= 80$ meV and $\bar{\omega}_{\rm Ru}^{\rm bulk} = $25 meV 
\cite{heid00,kim01}, and allow a 50\% variation of these values at the
surface, to plot eq. (\ref{freqopoor}) in Fig. \ref{frequency}
in the temperature range of interest to our study. As these particular
values for the characteristic frequencies are not well justified,
but should only be considered as rough estimates, we also include in the
graph the corresponding $\gamma_{\rm O-poor}^{\rm vib.}(T,p)$, if these
values changed by $\pm 50$\%. From Fig. \ref{frequency} we see that
the vibrational contribution to the surface free energy stays within
10 meV/{\AA}${}^2$ in all of the considered cases, and that the
uncertainty in the characteristic frequencies translates primarily
to variations of $\gamma_{\rm O-poor}^{\rm vib.}(T,p)$ at low
temperatures, where the value of the latter is very small anyway.

We have also computed the vibrational contribution to the surface free
energy of non-stoichiometric terminations in an analogous manner. There, 
the formulas turn out more lengthy, cf. eq. (\ref{gammamrich}), and
the vibrational contribution includes not only differences between
bulk and surface vibrational modes, but also absolute $F^{\rm vib.}(T,\omega)$
terms due to the excess or deficient atoms. Yet, even then the
vibrational contribution stays within $\pm 10$ meV/{\AA}${}^2$ similar
to the above described stoichiometric case. In conclusion, we therefore
take this value to represent a good upper bound for the vibrational
influence on the surface free energy.

Such a $\pm 10$ meV/{\AA}${}^2$ contribution is certainly not a completely
negligible factor, yet as we will show below it is of the same order as the
numerical uncertainty in our calculations. Furthermore, as will become
apparent in the discussion of the results, this uncertainty does not
affect any of the physical conclusions drawn in the present application.
Hence, we will henceforth neglect the complete vibrational contribution
to the Gibbs free energy, leaving only the total energies 
$E^{\rm total}(V,N_{\rm Ru},N_{\rm O})$ as the predominant term. 
In turn, this allows us to rewrite eq. (\ref{gammamrich}) to

\begin{eqnarray}
\label{gammamrich2}
&&\gamma_{\rm O-poor}(T,p) \approx \frac{1}{2A}
\bigg[ \; E^{\rm slab}(V,N_{\rm Ru},N_{\rm O})
\;- \\ \nonumber
            && \left. \frac{N_{\rm O}}{2} E^{\rm bulk}_{\rm RuO_2}(V) -
               \left( N_{\rm Ru} - \frac{N_{\rm O}}{2} \right)
               E^{\rm bulk}_{\rm Ru}(V) \right],
\end{eqnarray}

\noindent
which now contains exclusively terms directly accessible to a DFT
calculation. We stress, that this approximation is well justified
in the present case, but it is not a general result: There might
well be applications, where the inclusion of vibrational effects
on the surface free energy can be crucial.

\subsection{Pressure and temperature dependence of \boldmath$\mu_{\rm O}(T,p)$}

Having completely described the recipe of how to obtain $\gamma(T,p)$
as a function of the O chemical potential, the remaining task is to
relate the latter to a given temperature, $T$, and pressure, $p$. As
the surrounding O${}_2$ atmosphere forms an ideal gas like reservoir,
we can obtain in the appendix the following expression

\begin{equation}
\label{Ochem}
\mu_{\rm O}(T,p) = \mu_{\rm O}(T,p^{\circ}) \;+\; 1/2 \;k T
 \;ln \left( \frac{p}{p^{\circ}} \right),
\end{equation}

\noindent
which already gives us the temperature and pressure dependence, if
we only know the temperature dependence of $\mu_{\rm O}(T,p^{\circ})$
at one particular pressure, $p^{\circ}$.

\begin{table}
\caption{\label{tableI}
$\mu_{\rm O}(T,p^{\circ})$ in the temperature range of interest
to our study. The entropy and enthalpy changes used to obtain
$\mu_{\rm O}(T,p^{\circ})$ via eq. (\ref{entrdiff}) are taken
from the JANAF thermochemical tables at
$p^{\circ} = 1 {\rm atm}$ \cite{JANAF}.}

\begin{tabular}{rr | rr}
$T$   &  $\mu_{\rm O}(T,p^{\circ})$ & $T$   &  $\mu_{\rm O}(T,p^{\circ})$ \\ \hline
100 K &  -0.08 eV  & 600 K  & -0.61 eV \\ 
200 K &  -0.17 eV  & 700 K  & -0.73 eV \\
300 K &  -0.27 eV  & 800 K  & -0.85 eV \\
400 K &  -0.38 eV  & 900 K  & -0.98 eV \\
500 K &  -0.50 eV  &1000 K  & -1.10 eV \\
\end{tabular}
\end{table}

We choose as the zero reference state of $\mu_{\rm O}(T,p)$ the
total energy of oxygen in an isolated molecule, i.e. $\mu_{\rm O}(0 {\rm K},p) = 1/2
\; E^{\rm total}_{\rm O_2} \equiv 0$. With respect to this zero, reached
at our theoretical oxygen-rich limit, cf. eq. (\ref{oxygenlimits}),
$\mu_{\rm O}(T,p^{\circ})$ is then given by

\begin{eqnarray}
\label{entrdiff}
\mu_{\rm O}(T,p^{\circ}) &=& \mu_{\rm O}^{\rm O-rich}(0 {\rm K},p^{\circ})
                           \;+\; 1/2 \; \Delta G(\Delta T,p^{\circ},{\rm O_2}) \;= \nonumber \\ 
&&                               1/2 \;\left( H(T,p^{\circ},{\rm O_2}) - H(0{\rm K},p^{\circ},{\rm O_2}) \right) \; -
\nonumber \\ 
&&                            1/2 \; T \;\left( S(T,p^{\circ},{\rm O_2}) - S(0{\rm K},p^{\circ},{\rm O_2}) \right),
\end{eqnarray} 

\noindent
where we have used the relation $G = H - TS$ between the Gibbs
free energy and the enthalphy, $H$. This allows us to obtain the
aspired temperature dependence simply from the differences in
the enthalpy and entropy of an O${}_2$ molecule with respect to
the $T = 0$ K limit. For standard pressure, $p^{\circ} =$ 1 atm,
these values are e.g. tabulated in the JANAF thermochemical tables
\cite{JANAF}. Plugging them into eq. (\ref{entrdiff}) leads finally
to $\mu_{\rm O}(T,p^{\circ})$, which we list in Table
\ref{tableI}.

Together with eq. (\ref{Ochem}) the O chemical potential can thus
be obtained for any given $(T,p)$ pair. Although we prefer to
conveniently present the resulting surface energies as a 
one-dimensional function of $\mu_{\rm O}(T,p)$, we will often
convert this dependence also into a temperature (pressure) dependence
at a fixed pressure (temperature) in a second $x$-axis to
elucidate the physical meaning behind the obtained curves.

\subsection{DFT basis set and convergence}

The DFT input to eq. (\ref{gammamrich2}) has been obtained using the
full-potential linear augmented plane wave method (FP-LAPW) 
\cite{blaha99,kohler96,petersen00} within the generalized gradient
approximation (GGA) of the exchange-correlation functional \cite{perdew96}.
For the RuO${}_2$(110) surface calculation we use a symmetric
slab consisting of 3 rutile O-(RuO)-O trilayers, where all atomic
positions within the outermost trilayer were fully relaxed. A vacuum
region of $\approx$ 11 {\AA} is employed to decouple the surfaces of
consecutive slabs in the supercell approach. Test calculations with
5 and 7 trilayered slabs, as well as with a vacuum region up to
$\approx$ 28 {\AA} confirmed the good convergence of this chosen setup
with variations of $\gamma(T,p)$ smaller than $\pm$ 3 meV/{\AA}${}^2$.
Allowing a relaxation of deeper surface layers in the thicker slabs
did not result in a significant variation of the respective atomic
positions, nor did it influence the near-surface geometry as obtained
in the calculations with the standard 3 trilayer slabs. To ensure
maximum consistency, the corresponding RuO${}_2$ bulk computations are
done in exactly the same (110) oriented unit cell as used for the
slabs, in which the prior vacuum region is simply replaced by
additional RuO${}_2$ trilayers.

The FP-LAPW basis set is taken as follows: $R_{\rm{MT}}^{\rm{Ru}}=$1.8 bohr,
$R_{\rm{MT}}^{\rm{O}}=$1.3 bohr, wave function expansion inside the muffin
tins up to $l_{\rm{max}}^{\rm{wf}} = 12$, potential expansion up to
$l_{\rm{max}}^{\rm{pot}} = 4$. For the RuO${}_2$(110) slabs the Brillouin
zone integration was performed using a $(5 \times 10 \times 1)$ Monkhorst-Pack
grid with 15 {\bf k}-points in the irreducible part. The energy cutoff for
the plane wave representation in the interstitial region between the muffin
tin spheres was 17 Ry for the wave functions and 169 Ry for the potential.
Checking on the convergence, the surface free energies of the three possible
$(1 \times 1)$ RuO${}_2$(110) truncations discussed below were found
unchanged to within
1 meV/{\AA}${}^2$ by increasing the k-mesh to a $(7 \times 14 \times 1)$
Monkhorst-Pack grid with 28 {\bf k}-points in the irreducible part. A
larger interstitial cutoff up to 24 Ry reduced the absolute values of
the three $\gamma(T,p)$ by up to 10 meV/{\AA}${}^2$, yet as all of them
were reduced alike, their respective differences (which are the only
relevant quantities entering the physical argument) stayed constant
to within 5 meV/{\AA}${}^2$.

In total we thus find the numerical accuracy of the calculated surface
free energies with respect to the supercell approach and the finite basis
set to be within 10 meV/{\AA}${}^2$, which will not affect any of
the physical conclusions drawn. Note, that the stated imprecision does
not include possible errors introduced by more general deficiencies of
the approach, namely the use of the GGA as exchange-correlation functional,
on which we will comment below.

\section{Results}

\subsection{\boldmath RuO${}_2$(110) surface structure}

\begin{figure}
       \epsfxsize=0.35\textwidth \centerline{\epsfbox{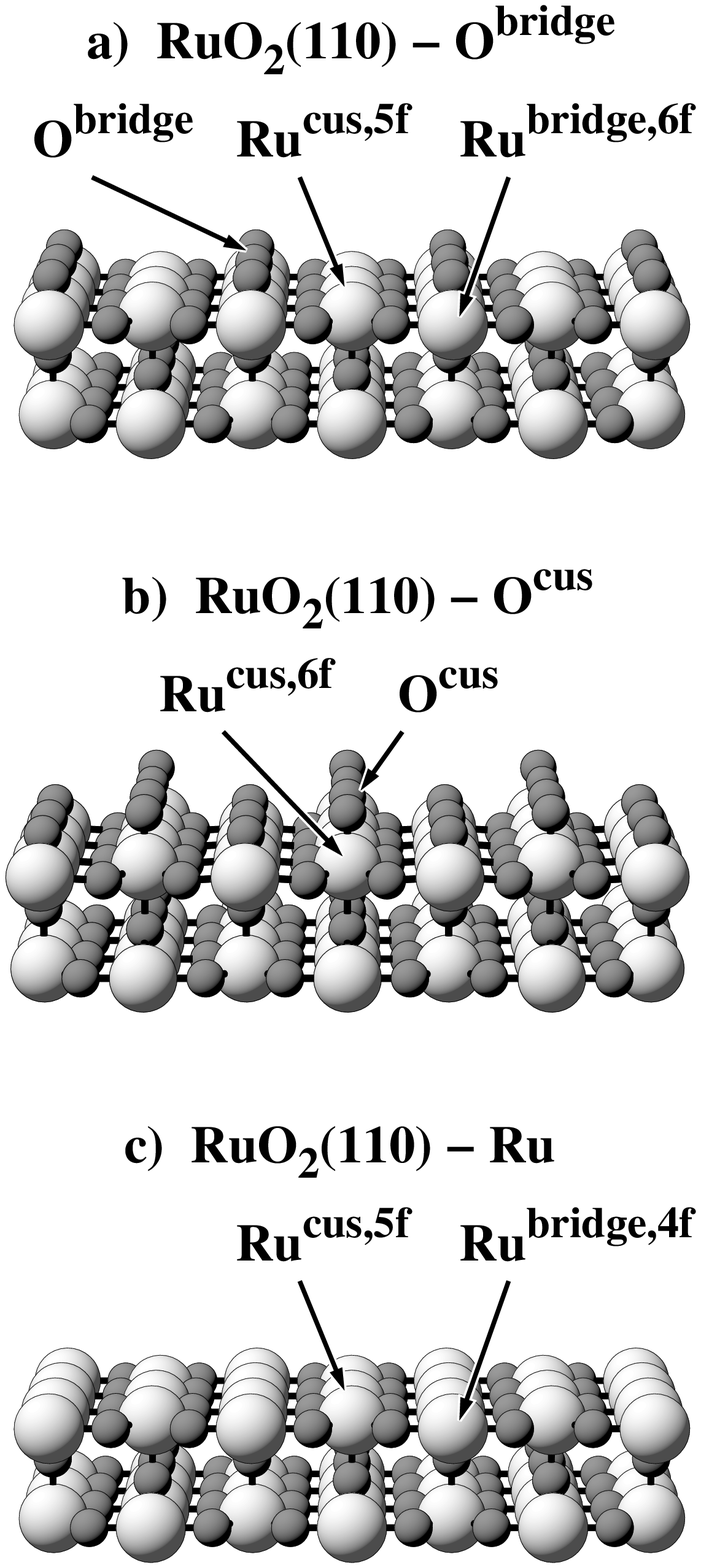}}
       \caption{Three possible terminating planes of the rutile (110)
        surface: a) Stoichiometric RuO${}_2$(110)-O${}^{{\rm bridge}}$
        termination with five-, six- and twofold coordinated
        Ru${}^{{\rm cus, 5f}}$, Ru${}^{{\rm bridge, 6f}}$ and O${}^{{\rm bridge}}$
        atoms respectively. b) RuO${}_2$(110)-O${}^{{\rm cus}}$
        termination, where additional O${}^{{\rm cus}}$ atoms sit atop the
        formerly undercoordinated Ru${}^{{\rm cus, 6f}}$ atoms. c)
        RuO${}_2$(110)-Ru termination, which lacks the O${}^{{\rm bridge}}$
        atoms in comparison to the stoichiometric termination (Ru = large,
        light spheres, O = small, dark spheres).
        \label{surfstructures}}
\end{figure}

RuO${}_2$ crystallizes in the rutile bulk structure, in which every metal
atom is coordinated to six oxygens, and every oxygen to three metal
neighbors \cite{sorantin92}. The oxygens forming an octahedron around
each Ru atom are not all equivalent, but can be distinguished into four
basal and two apical O atoms with calculated O-Ru bondlengths of 2.00 {\AA}
and 1.96 {\AA} respectively. We notice that along the (110) direction
this structure can then be viewed as a stacking sequence of O-(RuO)-O
trilayers, in which each trilayer is simply composed of an alternating
sequence of in-plane and up-right oriented oxygen-ruthenium coordination
octahedra, cf. Fig. \ref{surfstructures}b. Cut along (110), the rutile bulk
structure can therefore exhibit three distinct terminations of $(1 \times 1)$
periodicity, depending at which plane the trilayer is truncated, cf. Fig.
\ref{surfstructures}a-c.

Traditionally, the stoichiometric RuO${}_2$(110)-O${}^{\rm bridge}$
termination is believed to be the most stable one for all (110) surfaces
of rutile-structured crystals \cite{henrich94,noguera96}, as in the ionic
model it leads to an uncharged surface and cuts the least number of bonds: While
the Ru${}^{\rm bridge,6f}$ atoms possess their ideal sixfold O coordination
with two of their basal oxygens forming the terminal O${}^{\rm bridge}$
atoms, only the Ru${}^{\rm  cus, 5f}$ lack one apical on-top O, cf. Fig.
\ref{surfstructures}a. Note, that we will use a nomenclature for the surface
Ru atoms, where apart from a site specific characterization (e.g. cus for 
the coordinatively unsaturated site in the stoichiometric termination)
also the number of direct O neighbors (e.g. 5f for 5-fold coordination)
is stated. Vice versa, we indicate for the surface O atoms to which specific
site they bind (e.g. O${}^{{\rm bridge}}$ binds to the Ru${}^{{\rm bridge, 
6f}}$ atoms).

Alternatively, in the second possible RuO${}_2$(110)-O${}^{\rm cus}$ 
termination, cf. Fig. \ref{surfstructures}b, terminal O${}^{\rm cus}$
atoms occupy sites atop of the formerly undercoordinated Ru${}^{\rm cus,
6f}$ atoms, so that now all metal atoms in the surface possess their
ideal sixfold coordination. This is, of course, paid for by the presence
of both the only twofold and onefold coordinated O${}^{\rm bridge}$ and
O${}^{\rm cus}$ atoms, respectively. Finally, the third RuO${}_2$(110)-Ru
termination exhibiting the mixed (RuO) plane at the surface is achieved by
removing the O${}^{\rm bridge}$ atoms from the stoichiometric termination,
cf. Fig. \ref{surfstructures}c. Here, no undercoordinated oxygens are present
any more, yet, at the expense of the four- and fivefold bonded Ru${}^{\rm
bridge,4f}$ and Ru${}^{\rm cus,5f}$ atoms.

\subsection{Prediction of a high pressure termination}

\begin{figure}
       \epsfxsize=0.49\textwidth \centerline{\epsfbox{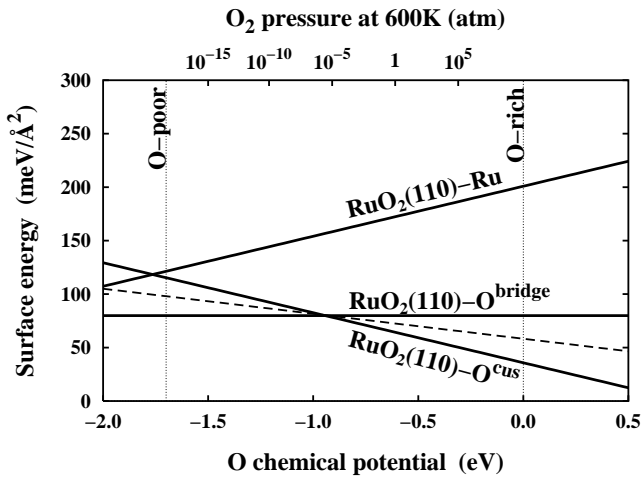}}
       \caption{Surface free energies, $\gamma(T,p)$, of the three
        RuO${}_2$(110) terminations depicted in Fig. \ref{surfstructures}.
        Additionally shown as dashed line is the surface free energy
        of a RuO${}_2$(110)-O${}^{\rm cus}$ termination, in which only
        every second O${}^{\rm cus}$ site along the trenches is occupied.
        The dotted vertical lines indicate the allowed range of the
        oxygen chemical potential, $\mu_{\rm O}(T,p)$, using $1/2\; E^{\rm
        total}_{\rm O_2}$ as zero reference as explained in Section
        IIB. In the top $x$-axis, the dependence on $\mu_{\rm O}(T,p)$
        has been cast into a pressure scale at a fixed
        temperature of $T=$ 600 K.
        \label{surfenergy1}}
\end{figure}

The calculated surface free energies of the three possible terminations
are shown in Fig. \ref{surfenergy1}. As explained in connection with
eq. (\ref{gammaorich}), the RuO${}_2$(110)-O${}^{\rm cus}$
(RuO${}_2$(110)-Ru) termination with an excess (deficiency) of O at
the surface becomes more and more favorable (unfavorable) towards the
O-rich limit, while the stoichiometric RuO${}_2$(110)-O${}^{\rm bridge}$
termination exhibits a constant $\gamma(T,p)$. Indeed, we find the
traditionally assumed stoichiometric RuO${}_2$(110)-O${}^{\rm bridge}$
surface to be the most stable over quite a range of oxygen chemical
potentials above the O-poor limit. To make this range a bit more lucid,
we have used eq. (\ref{Ochem}) to convert $\mu_{\rm O}(T,p)$ into a
physically more comprehensible pressure dependence at a fixed temperature,
cf. the top $x$-axis of Fig. \ref{surfenergy1}. The temperature of
$T = 600$ K corresponds to a typical annealing temperature employed
experimentally for this system
\cite{kim01,atanasoska88,boettcher99a,over00,kim01b}. From the
resulting pressure scale we see, that the stability of the stoichiometric
termination extends therefore roughly around the pressure range
corresponding to UHV conditions.

Yet, this is different at higher O pressures, where the
RuO${}_2$(110)-O${}^{\rm cus}$ termination becomes the most stable
surface structure, cf. Fig. \ref{surfenergy1}. In the O-rich limit,
it exhibits a $\gamma^{\rm O-rich}(T,p)$, which is by 49
meV/{\AA}${}^2$ lower than the one of the stoichiometric
RuO${}_2$(110)-O${}^{\rm bridge}$ surface, i.e. the deduced crossover
between the two terminations is far beyond the estimated uncertainty
of $\approx \pm 10$ meV/{\AA}$^{2}$ due to the neglection of the
vibrational contribution to the Gibbs free energies and due to the
finite basis set. This estimate does, however, not comprise the
more general error due to the use of the GGA as exchange-correlation
functional. To this end, we have also calculated the surface free
energies of the two competing terminations within the local density
approximation (LDA) \cite{perdew92}. Although the absolute values
of both $\gamma^{\rm O-poor}(T,p)$ turn out to be by $\approx$
15 meV/{\AA}${}^2$ higher, their respective difference is almost
unchanged, which is eventually what determines the crossover point
of the two lines in Fig. \ref{surfenergy1}. The 
RuO${}_2$(110)-O${}^{\rm cus}$ termination is therefore the lowest
energy structure for $\mu_{\rm O}(T,p) > -0.85$ eV in the LDA, which
is almost the same as the $\mu_{\rm O}(T,p) > -0.93$ eV found with the
GGA, shown in Fig. \ref{surfenergy1}. Consequently, while the choice
of the exchange-correlation functional may affect the exact transition
temperature or pressure, the transition {\em per se} is untouched.
In turn, we may safely predict the stability of a polar surface
termination on RuO${}_2$(110) at high O chemical potential, corresponding
e.g. to the pressure range typical for catalytic applications. 

Note, that Fig. \ref{surfenergy1} summarizes only the $\gamma(T,p)$
of the three $(1 \times 1)$ terminations, which arise by truncating the
RuO${}_2$ crystal at bulk-like planes in the (110) orientation. Yet, it
is {\em a priori} not clear, that the terminal atoms at the surface
must be in the sites corresponding to the bulk stacking sequence.
To check this, we have additionally calculated the surface free
energies of surface structures, where the O${}^{\rm bridge}$
(O${}^{\rm cus}$) atoms occupy atop (bridge) sites over the
Ru${}^{\rm bridge,5f}$ (Ru${}^{\rm cus,6f}$) atoms, instead of
their normal bridging (atop) configuration. In both cases, we
find the $\gamma(T,p)$ considerably higher, which excludes the
possibility that these adatoms occupy non bulk-like sites at
the surface. Similarly, a RuO${}_2$(110)-O${}^{\rm cus}$ termination,
where the O${}^{\rm bridge}$ atoms have been removed, can also safely
be ruled out as alternative for a stoichiometrically terminated
surface.

\subsection{Lateral interaction between \boldmath O${}^{\rm cus}$ atoms and vacancy concentration}

The reasoning of the last Section leaves in fact only the
RuO${}_2$(110)-O${}^{\rm bridge}$ and RuO${}_2$(110)-O${}^{\rm cus}$
termination as the relevant surface structures stabilized in UHV
and under high O pressure respectively. Both differ from each other
only by the presence of the additional O${}^{\rm cus}$ atoms, which
continue the bulk stacking sequence by filling the vacant sites
atop of the formerly undercoordinated Ru${}^{\rm cus,5f}$ atoms.
The way the RuO${}_2$(110)-O${}^{\rm cus}$ termination is formed at
an increasing oxygen chemical potential will therefore depend
significantly on the details of the lateral interaction among
this adatom species. 

The RuO${}_2$(110)-O${}^{\rm bridge}$ surface has a trench-like
structure with a distance of 6.43 {\AA} between the rows formed
by the O${}^{\rm bridge}$ atoms, cf. Fig. \ref{surfstructures}a.
This renders any lateral interaction between O${}^{\rm cus}$ atoms
adsorbed in neighboring trenches rather unlikely. On the other
hand, the distance between two O${}^{\rm cus}$ atoms occupying
neighboring sites along one trench is only 3.12 {\AA}. To check
on the corresponding interaction we calculated the surface free
energy of a RuO${}_2$(110)-O${}^{\rm cus}$ termination in a
$(2 \times 1)$ supercell, in which the O${}^{\rm cus}$ atoms
occupied only every other site along the trenches. The
corresponding $\gamma(T,p)$ is drawn as a dashed line in Fig.
\ref{surfenergy1}. As now only half of the excess O${}^{\rm cus}$
atoms are present, the slope of this curve has to be one half
of the slope of the line representing the normal 
RuO${}_2$(110)-O${}^{\rm cus}$ termination, cf. eq. (\ref{gammaorich}).

Interestingly, both curves cross the stoichiometric
RuO${}_2$(110)-O${}^{\rm bridge}$ line in exactly the same point.
This can only be understood by assuming a negligible lateral
interaction between neighboring O${}^{\rm cus}$ atoms: If there
was an attractive (repulsive) interaction between them, then
it would be favorable (unfavorable) to put O${}^{\rm cus}$
atoms as close to each other as possible. In turn, the $(2 \times 1)$
overlayer of O${}^{\rm cus}$ atoms, in which only every other
site is occupied, would be less (more) stable than the normal
RuO${}_2$(110)-O${}^{\rm cus}$ termination, where all neighboring
sites are full. Consequently, the stability with respect to the
stoichiometric termination would be decreased (enhanced), leading
to a later (earlier) crossover point in Fig. \ref{surfenergy1}.
That both calculated lines cross the RuO${}_2$(110)-O${}^{\rm bridge}$
line at the same point is therefore a reflection of a negligible
lateral interaction between the O${}^{\rm cus}$ atoms.

Additionally, we compute a very high barrier of almost 1.5 eV
for diffusion of O${}^{\rm cus}$ atoms along the trenches,
indicating that the latter species will be practically immobile
in the temperature range, where the oxide is stable. This
together with the small lateral interaction indicates that
at increasing O chemical potential the RuO${}_2$(110)-O${}^{\rm cus}$
surface is formed from the stoichiometric termination by a
random occupation of O${}^{\rm cus}$ sites, until eventually
the whole surface is covered. Not withstanding, at finite temperatures
there will still be a certain vacancy concentration even
at O chemical potentials above the crossover point of the two
terminations. As the undercoordinated Ru${}^{\rm cus,5f}$
atoms exposed at such a vacant O${}^{\rm cus}$ site, cf. Fig.
\ref{surfstructures}, might be chemically active sites for
surface reactions \cite{over00}, it is interesting to
estimate how many of these sites will be present under given
$(T,p)$ conditions.

Since we have shown that each O${}^{\rm cus}$ site at the
surface is filled independently from the others, we can
calculate its occupation probability within a simple two level
system (site occupied or vacant) in contact with a heat
bath. The vacancy concentration follows then from a canonic
distribution, where the energy of the two levels is given
by the $\gamma(T,p)$ of the two terminations at the
chosen chemical potential. As an example, we first address
room temperature, where the RuO${}_2$(110)-O${}^{\rm cus}$
termination becomes stable at pressures higher than $p \sim 10^{-22}$
atm. A vacancy concentration of only $1\%$ is in turn already
reached at $p \sim 10^{-17}$ atm, so that at this temperature
there will only be a negligible number of vacancies on the
O${}^{\rm cus}$-terminated surface for any realistic pressure.

However, this situation becomes completely different at
elevated temperatures. At $T = 800$ K, the crossover to
the RuO${}_2$(110)-O${}^{\rm cus}$-termination occurs at
$p \sim 10^{-1}$ atm with a $10\%$ vacancy concentration
still present at 10${}^{2}$ atm. In the range of atmospheric
pressures the RuO${}_2$(110) surface will therefore exhibit a
considerable number of vacancies, which could explain
the high catalytic activity reported for this material
\cite{over00,kim01b,boettcher97,boettcher00,fan01,wang01}.
However, although we have deliberately chosen $T=800$ K
as a typical catalytic temperature, where e.g. a maximum
conversion rate for the CO/CO${}_2$ oxidation reaction
over RuO${}_2$(110) was found \cite{boettcher00}, we 
immediately stress that our reasoning is at the moment
only based on the O pressure alone and therefore not
directly applicable to catalysis experiments, which may
also depend on the partial pressure of other reactants
in the gas phase.

\subsection{On the stability of polar surfaces}

As already mentioned in the introduction, the predicted high
pressure RuO${}_2$(110)-O${}^{\rm cus}$ termination is
traditionally not expected as it forms a so-called
polar surface, which should not be stable on electrostatic grounds
\cite{henrich94,noguera96}. The corresponding argument is based on the
ionic model of oxides, in which every atom in the solid is assumed to
be in its bulk formal oxidation state. Along a particular direction, $z$,
the crystal may then be viewed as a stack of planes with charge $q$,
each of which contribute with $V(z) \propto qz$ to the total
electrostatic potential. As this contribution diverges at infinite
distances the crystal as a whole can in turn only be stable, if
constructed as a neutral block, in which all infinities due to the
individual planes cancel. For RuO${}_2$(110), which is a type 2
surface in Tasker's widely used classification scheme \cite{tasker79}, 
the only neutral repeat unit is a symmetric O-(RuO)-O trilayer with a
(-2)-(+4)-(-2) charge sequence, so that the only surface termination
without net dipole moment would correspondingly be the stoichiometric
RuO${}_2$(110)-O${}^{\rm bridge}$ termination.

\begin{figure}
       \epsfxsize=0.49\textwidth \centerline{\epsfbox{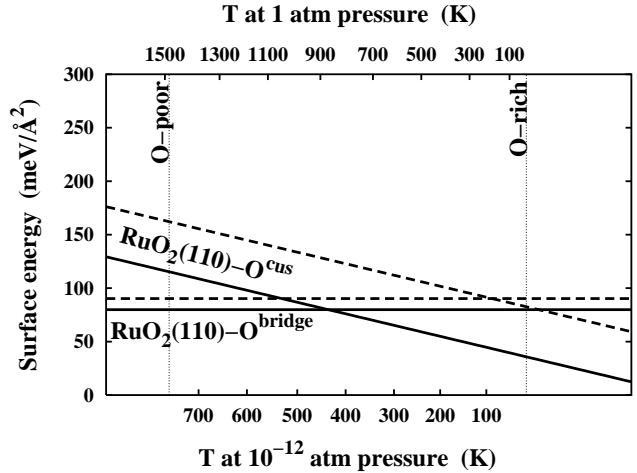}}
       \caption{Surface free energies, $\gamma(T,p)$, of the
        stoichiometric RuO${}_2$(110)-O${}^{\rm bridge}$ and polar
        RuO${}_2$(110)-O${}^{\rm cus}$ termination. Shown is the
        effect of relaxation at the surface, with solid lines indicating
        fully relaxed surface structures and dashed lines the
        corresponding bulk-truncated geometries. The dependence on
        the oxygen chemical potential has been translated into a
        temperature scale at $10^{-12}$ atm (bottom $x$-axis) and
        1 atm (top $x$-axis) pressure.
        \label{surfenergy2}}
\end{figure}

On the other hand, the RuO${}_2$(110)-O${}^{\rm cus}$ termination with
its extra unmatched (-2) charge plane formed by the O${}^{\rm cus}$ atoms
would lead to a diverging potential and should thus not be stable. That
we indeed find this surface stabilized at higher O chemical potentials
points to the most obvious shortcoming of the electrostatic model, namely
the assumption that all atoms in the solid are identical, i.e. that 
also all surface atoms are both structurally as well as electronically
in a bulk-like state. In how much already the additional structural
degrees of freedom at the surface influence the stability is exemplified
in Fig. \ref{surfenergy2}, where the surface free energies of the two
relevant terminations are compared in either a bulk-truncated or the
fully relaxed geometry. While the small relaxation of the
RuO${}_2$(110)-O${}^{\rm bridge}$ termination hardly affects the
$\gamma(T,p)$, the bulk-truncated RuO${}_2$(110)-O${}^{\rm cus}$
surface turns out considerably less stable compared to its relaxed counterpart.
The by +2.5 eV strongly increased workfunction of the bulk-truncated
RuO${}_2$(110)-O${}^{\rm cus}$ surface with respect to the stoichiometric
termination shows that the addition of the O${}^{\rm cus}$ atoms indeed
induces a considerable dipole moment as suggested by the ionic model.
Yet, just the relaxation lowers this workfunction again by 1 eV,
indicating that the dipole moment can already be considerably reduced via
a strongly shortened O${}^{\rm cus}$-Ru${}^{\rm cus,6f}$ bondlength
of 1.70 {\AA} (compared to the bulk-like 1.96 {\AA}), therewith considerably
stabilizing the surface.

\begin{figure}
       \epsfxsize=0.40\textwidth \centerline{\epsfbox{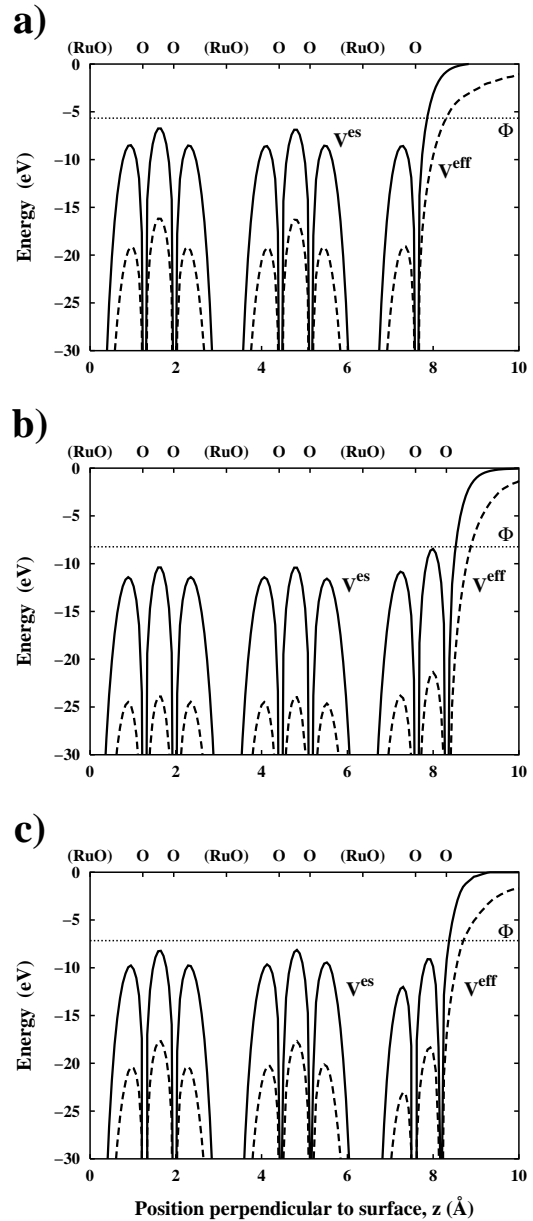}}
       \caption{($x,y$)-averaged Kohn-Sham effective, $V^{\rm eff}(z)$ (dashed
        line), and electrostatic potential, $V^{\rm es}(z)$ (solid line),
        along the (110) direction perpendicular to the surface, $z$. Additionally
        shown is the workfunction, $\Phi$, as dotted line.
        a) Bulk-truncated RuO${}_2$(110)-O${}^{\rm bridge}$ termination
        ($\Phi = 5.7$ eV), b) Bulk-truncated RuO${}_2$(110)-O${}^{\rm cus}$
        termination ($\Phi = 8.2$ eV), and c) Fully relaxed
        RuO${}_2$(110)-O${}^{\rm cus}$ termination ($\Phi = 7.2$ eV).
        The top $x$-axis marks the position of O-(RuO)-O layers in the
        crystal. 
        \label{potential}}
\end{figure}

Yet, not only structurally, but also electronically the topmost
layers in the RuO${}_2$(110)-O${}^{\rm cus}$ surface differ
appreciably from their respective bulk counterparts. This is
illustrated in Fig. \ref{potential}, where we show the ($x,y$)-averaged
potential along the (110) direction perpendicular to the surface, $z$.
In the bulk-truncated, stoichiometric RuO${}_2$(110)-O${}^{\rm bridge}$ 
termination in Fig. \ref{potential}a the electrostatic potential at
the topmost O-(RuO)-O trilayer is still almost identical to the
corresponding one in the deeper trilayers, thus enabling in this
case a description of this surface in terms of bulk-like planes
as assumed in the ionic model. On the contrary, we find a significant
deviation of the potential in the outermost layers of the
RuO${}_2$(110)-O${}^{\rm cus}$ termination, cf. Fig. \ref{potential}b,
even for a bulk-truncated geometry. This difference is further
enhanced by the structural relaxation, cf. Fig. \ref{potential}c,
so that the topmost (RuO)-O-O planes of this termination are
certainly not well characterized by bulk properties, thus
invalidating the electrostatic argument raised against this
polar surface.

Instead, we argue that the surface fringe composed by the topmost
layers should be viewed as a new material, which properties might
differ considerably from the bulk stacking sequence due to the
additional structural and electronic degrees of freedom present
at the surface. A similar conclusion has previously been reached
by Wang {\em et al.} \cite{wang98,wang00}, who discussed the
stability of oxygen-terminated polar (0001) surfaces of
corundum-structured $\alpha$-Fe${}_2$O${}_3$ and $\alpha$-Al${}_2$O${}_3$.
This indicates, that the traditionally dismissed polar terminations
\cite{noguera00} might indeed be a more general phenomenon, which
existence could be a crucial ingredient to understand the function
of oxide surfaces under realistic environmental conditions. As
particularly polar terminations with excess oxygen can be stabilized at
increased O${}_2$ partial pressure in the gas phase, the different
properties of the latter should be taken into consideration, when
modelling high pressure applications like catalysis.

\subsection{Importance of experimental preparation conditions}

This influence of the O${}_2$ partial pressure on the
surface morphology and function has recently become apparent
in a number of studies addressing the reported high CO oxidation
rates over Ru catalysts. While it was for a long time believed
that the active species is the Ru metal itself
\cite{madey75,cant78,peden86,stampfl96}, the decisive role
played by oxide patches formed under catalytic conditions was
only recently realized
\cite{boettcher99a,over00,kim01b,boettcher97,boettcher00}.
This was primarily due to the problem of preparing a fully
oxidized surface in a controlled manner under UHV conditions.
Yet, by means of more oxidizing carrier gases or higher O partial
pressures it is now possible to circumvent this {\em materials gap}
\cite{boettcher99b}, enabling the detailed characterization
of RuO${}_2$(110) domains formed on the model Ru(0001) surface
with the techniques of surface science
\cite{kim01,boettcher99a,over00,kim01b,boettcher97,boettcher00,fan01,wang01}.

However, the results of the present study show that the surface
termination of these domains changes with the O${}_2$ pressure,
too. Not aware of this dependence a reaction mechanism for
the CO oxidation was previously proposed, situated solely on
the stoichiometric RuO${}_2$(110)-O${}^{\rm bridge}$ termination
that was characterized in the respective LEED study under UHV
conditions \cite{over00,kim01b}. While it is presently not clear,
to what extent the O${}^{\rm cus}$ atoms additionally present
under catalytic pressures are involved in the reaction
\cite{fan01,wang01}, this still highlights the delicacy with which the
results of UHV spectroscopies and post-exposure experiments have
to be applied to model catalysis and steady-state conditions.

Only very recently UHV studies achieved to stabilize the high-pressure
RuO${}_2$(110)-O${}^{\rm cus}$ termination intentionally
by postdosing O${}_2$ at low temperatures \cite{kim01,boettcher99a,fan01,wang01}.
Temperature desorption spectroscopy (TDS) experiments found the
corresponding excess O${}^{\rm cus}$
atoms to be stable up to about 300-550 K in UHV \cite{boettcher99a}.
This agrees nicely with the calculated transition temperature
of $450 \pm 50$ K at the crossover point between the two
terminations for a pressure of $10^{-12 \pm 2}$ atm, presumably
present during a typical TDS experiment \cite{equilibrium},
cf. the bottom $x$-axis of Fig. \ref{surfenergy2}. Yet, the actual
desorption temperature is of course significantly higher for the
orders of magnitude higher O${}_2$ partial pressures present
in catalytic applications. This is exemplified by the temperature
scale on the top $x$-axis of Fig. \ref{surfenergy2} at a pressure
of $p = 1$ atm, representative for the early high-pressure
experiments addressing the high CO/CO${}_2$ conversion rates of
Ru catalysts \cite{cant78,peden86}. The corresponding elevated
transition temperature of 900 K shows that O${}^{\rm cus}$ atoms
were most probably present on oxidized RuO${}_2$(110) domains
in all of these experiments.

The presence of the hitherto unaccounted high-pressure termination
of RuO${}_2$(110) might therefore be an essential keystone to
understand the data obtained from grown RuO${}_2$(110) films or
oxidized Ru(0001) surfaces, which are unanimously prepared under
highly O-rich conditions. We already exemplified this in a
preceding publication \cite{reuter01} by suggesting that the long
discussed, largely shifted satellite peak in Ru $3d$ x-ray
photoemission spectroscopy (XPS) data
from such surfaces \cite{atanasoska88,kim97}, is due to the
Ru${}^{\rm cus,6f}$ atoms in the RuO${}_2$(110)-O${}^{\rm cus}$
termination, which experience a significantly different
environment due to the aforementioned, very short bondlength
to the O${}^{\rm cus}$ atoms.

If the high-pressure termination created during the preparation
of the crystal is frozen in during the transfer to UHV depends
then on the details of the transfer itself, e.g. on such
nitty-gritties of whether or not the temperature is kept constant
while the pressure goes down to its base value after exposure.
A dependence of the TDS data of an oxygen-rich Ru(0001) surface
on these parameters has already been reported and considered
by B\"ottcher and Niehus \cite{boettcher99a}, while the final
annealing step to 600 K after transfer to UHV employed in the
LEED work identifying RuO${}_2$(110) domains on oxidized Ru(0001)
\cite{over00,kim01b} explains, why there only the stoichiometric
RuO${}_2$(110)-O${}^{\rm bridge}$ termination could be characterized,
cf. Fig. \ref{surfenergy1}.

Such dependences on the experimental preparation have hitherto
often been neglected, entailing only a low comparability
of data sets obtained in different groups. Instead, the present
results demonstrate that systematic investigations in the whole
$(T,p)$ range are required to fully identify the surface structure
and composition of oxide surfaces at realistic conditions, which
in turn is a prerequisite before tackling the long-term goal
of understanding the function of the latter in the wealth of
everyday applications.

\section{Summary}

In conclusion we have combined density-functional theory (DFT)
and classical thermodynamics to determine the lowest
energy structure of an oxide surface in equilibrium with an O
environment. The formalism is applied to RuO${}_2$(110) showing that
apart from the expected stoichiometric surface, a so-called polar
termination with an excess of oxygen (O${}^{\rm cus}$) is stabilized
at high O chemical potentials. Depending on the details of the
experimental preparation conditions, either of the two terminations
can therefore be present, which different properties have to be
taken into account, when trying to understand the obtained data
or aiming to extrapolate the results of UHV {\em ex situ} techniques
to high-pressure applications like oxidation catalysis.

A polar termination is traditionally not conceived to be stable
within the framework of electrostatic arguments based on the ionic
model of oxides. We show that this reasoning is of little validity
as it assumes all atoms to be in the same bulk-like state. On the
contrary, the additional structural and electronic degrees of freedom
at an oxide surface allow substantial deviations from these bulk
properties and may thus stabilize even non-stoichiometric
surface terminations. A similar conclusion was previously reached
also for the O-rich (0001) termination of corundum-structured
$\alpha$-Fe${}_2$O${}_3$ indicating that polar surfaces might
indeed be a more general feature of transition metal oxides.
The concentration of oxygen vacancies found for the polar termination
of RuO${}_2$(110) at atmospheric pressures and elevated
temperatures could finally possibly explain the high catalytic
activity reported for this surface.

\begin{appendix}
\section{}

\subsection{Vibrational contribution to the Gibbs free energy}

The vibrational contribution to the Gibbs free energy comprises
vibrational energy and entropy, cf. eq. (\ref{ffreq}). Both
can be calculated from the partition function of an $N$-atomic
system \cite{ashcroft81}

\begin{equation}
Z \;=\; \sum_{i=1}^{3N} \int \frac{d{\bf k}}{(2\pi)^3} \;
\sum_{n = 0}^{\infty} e^{-(n+\frac{1}{2}) \beta \hbar \omega_i({\bf k})},
\end{equation}

\noindent
where $\beta = 1/kT$ and the $\omega_i({\bf k})$ are the $3N$
vibrational modes. The vibrational energy is then given by 

\begin{equation}
E^{\rm vib.}(T,V,N) \;=\; - \frac{\partial}{\partial \beta} ln Z,
\end{equation}

\noindent
and the entropy is defined as

\begin{equation}
S^{\rm vib.}(T,V,N) \;=\; k \left(\; ln Z \;+\; \beta E^{\rm vib.} \; \right).
\end{equation}

\noindent
Writing $F^{\rm vib.}(T,V,N)$ as a frequency integral including the
phononic density of states, $\sigma(\omega)$, and using the relation
$F^{\rm vib.} = E^{\rm vib.} - TS^{\rm vib.}$, one arrives at

\begin{eqnarray}
F^{\rm vib.}(T,\omega) &=& \hbar \omega \left( \frac{1}{2} \;+\;
     \frac{1}{e^{\beta \hbar \omega} - 1} \right) \;- \\ \nonumber
                       && kT \left[ \frac{\beta \hbar \omega}{e^{\beta \hbar \omega} -1}
\;-\; ln \left( 1 - e^{-\beta \hbar \omega} \right) \right]. 
\end{eqnarray}

\subsection{Ideal gas expression for \boldmath$\mu_{O_2}(T,p)$}

For an ideal gas of $N$ particles at constant pressure, $p$, and
temperature, $T$, the chemical potential is simply given by the Gibbs
free energy per atom,

\begin{equation}
\label{mug}
\mu \;=\; \left( \frac{\partial G}{\partial N} \right)_{T,p,N} \;=\; \frac{G}{N}.
\end{equation}

\noindent
As the Gibbs free energy is a potential function depending on pressure
and temperature, its total derivative can be written as

\begin{eqnarray}
dG &=& \left( \frac{\partial G}{\partial T} \right)_p dT \;+\;
       \left( \frac{\partial G}{\partial p} \right)_T dp \;= \\ \nonumber
   & & - S dT \;+\; V dp,
\end{eqnarray}

\noindent
where we have inserted the Maxwell relations for the entropy, $S$,
and volume, $V$. Using the ideal gas equation of state, $pV = NkT$,
the partial derivative of $G(T,p)$ with respect to pressure at constant
temperature is consequently

\begin{equation}
\left( \frac{\partial G}{\partial p} \right)_T \;=\; V \;=\; \frac{NkT}{p}.
\end{equation}

\noindent
In turn, a finite pressure change from $p$ to $p^{\circ}$ results in

\begin{eqnarray}
\label{gfinite}
G(T,p) - G(T,p^{\circ}) &=& \int_{p^{\circ}}^p \;
\left( \frac{\partial G}{\partial p} \right)_T dp \;= \\ \nonumber
                        & & NkT \; ln \left( p / p^{\circ} \right).
\end{eqnarray}

Combining eqs. (\ref{mug}) and (\ref{gfinite}), we can finally write
for the chemical potential of O

\begin{eqnarray}
\mu_{\rm O}(T,p) &=& 1/2 \;\mu_{\rm O_2}(T,p) \;= \\ \nonumber
                 &=& \mu_{\rm O}(T,p^{\circ}) \;+\;
                     1/2 \; k T \; ln \left( p / p^{\circ} \right),
\end{eqnarray}

\noindent
which is the expression used in Section IIE.

\end{appendix}


\begin{references}

\bibitem{stampfl01}
C. Stampfl, M.V. Ganduglia-Pirovano, K. Reuter, and M. Scheffler,
Surf. Sci. {\bf 500}, (accepted); preprint download under
http://www.fhi-berlin.mpg.de/th/pub01.html.

\bibitem{wang98}
X.-G. Wang, W. Weiss, Sh.K. Shaikhutdinov, M. Ritter, M. Petersen, F. Wagner,
R. Schl\"ogl, and M. Scheffler, Phys. Rev. Lett. {\bf 81}, 1038 (1998).

\bibitem{wang00}
X.-G. Wang, A. Chaka, and M. Scheffler, Phys. Rev. Lett. {\bf 84}, 3650 (2000).

\bibitem{henrich94}
V.E. Henrich and P.A. Cox, {\em The Surface Science of Metal Oxides}, Cambridge
University Press, Cambridge (1994).

\bibitem{noguera96}
C. Noguera, {\em Physics and Chemistry at Oxide Surfaces}, Cambridge
University Press, Cambridge (1996).

\bibitem{noguera00}
C. Noguera, J. Phys. Cond. Mat. {\bf 12}, R367 (2000).

\bibitem{CRC95}
{\em CRC Handbook of Chemistry and Physics}, 76th ed. (CRC Press, Boca Raton FL,
1995).

\bibitem{heid00}
R. Heid, L. Pintschovius, W. Reichardt, and K.-P. Bohnen,
Phys. Rev. B {\bf 61}, 12059 (2000).

\bibitem{kim01}
Y.D. Kim, A.P. Seitsonen, S. Wendt, J. Wang, C. Fan, K. Jacobi, H. Over,
and G. Ertl, J. Phys. Chem. {\bf 105}, 3752 (2001).

\bibitem{JANAF}
D.R. Stull and H. Prophet, {\em  JANAF Thermochemical Tables}, 2nd ed.,
U.S. National Bureau of Standards, Washington, D.C. (1971).

\bibitem{blaha99}
P. Blaha, K. Schwarz and J. Luitz, {\bf WIEN97}, \emph{A Full Potential
Linearized Augmented Plane Wave Package for Calculating Crystal
Properties}, Karlheinz Schwarz, Techn. Universit\"at Wien, Austria,
(1999). ISBN 3-9501031-0-4.

\bibitem{kohler96}
B. Kohler, S. Wilke, M. Scheffler, R. Kouba, and C. Ambrosch-Draxl,
Comput. Phys. Commun. {\bf 94}, 31 (1996).

\bibitem{petersen00}
M. Petersen, F. Wagner, L. Hufnagel, M. Scheffler, P. Blaha, and
K. Schwarz, Comp. Phys. Commun. {\bf 126}, 294 (2000).

\bibitem{perdew96}
J.P. Perdew, K. Burke and M. Ernzerhof, Phys. Rev. Lett. {\bf 77}, 3865 (1996).

\bibitem{sorantin92}
P.I. Sorantin and K.H. Schwarz, Inorg. Chem. {\bf 31}, 567 (1992).

\bibitem{atanasoska88}
Lj. Atanasoska, W.E. O'Grady, R.T. Atanasoski, and F.H. Pollak,
Surf. Sci. {\bf 202}, 142 (1988).

\bibitem{boettcher99a}
A. B\"ottcher and H. Niehus, Phys. Rev. B {\bf 60}, 14396 (1999).

\bibitem{over00}
H. Over, Y.D. Kim, A.P. Seitsonen, S. Wendt, E. Lundgren, M. Schmid,
P. Varga, A. Morgante, and G. Ertl, Science {\bf 287}, 1474 (2000).

\bibitem{kim01b}
Y.D. Kim, H. Over, G. Krabbes, and G. Ertl, Topics in Catalysis {\bf 14},
95 (2001).

\bibitem{perdew92}
J.P. Perdew and Y. Wang, Phys. Rev. B {\bf 45}, 13244 (1992).

\bibitem{boettcher97}
A. B\"ottcher, H. Niehus, S. Schwegmann, H. Over, and G. Ertl,
J. Phys. Chem. B {\bf 101}, 11185 (1997).

\bibitem{boettcher00}
A. B\"ottcher, H. Conrad, and H. Niehus, Surf. Sci. {\bf 452}, 125 (2000).

\bibitem{fan01}
C.Y. Fan, J. Wang, K. Jacobi, and G. Ertl,
J. Chem. Phys. {\bf 114}, 10058 (2001).

\bibitem{wang01}
J. Wang, C.Y. Fan, K. Jacobi, and G. Ertl, Surf. Sci. {\bf 481}, 113 (2001).

\bibitem{tasker79}
P.W. Tasker, J. Phys. C {\bf 12}, 4977 (1979).

\bibitem{madey75}
T.E. Madey, H.A. Engelhardt, and D. Menzel, Surf. Sci. {\bf 48}, 304 (1975).

\bibitem{cant78}
N.W. Cant, P.C. Hicks, and B.S. Lennon, J. Catal. {\bf 54}, 372 (1978).

\bibitem{peden86}
C.H.F. Peden and D.W. Goodman, J. Phys. Chem. {\bf 90}, 1360 (1986).

\bibitem{stampfl96}
C. Stampfl, S. Schwegmann, H. Over, M. Scheffler, and G. Ertl,
Phys. Rev. Lett. {\bf 77}, 3371 (1996).

\bibitem{boettcher99b}
A. B\"ottcher and H. Niehus, J. Chem. Phys. {\bf 110}, 3186 (1999).

\bibitem{equilibrium}
A TDS experiment with a low heating rate of 6 K/s is considered
to refer to a situation, where the system is always close to equilibrium.
Indeed, the agreement of the experimental results of B\"ottcher
and Niehus \cite{boettcher99b} with our calculations confirmes
this assumption.

\bibitem{reuter01}
K. Reuter and M. Scheffler, Surf. Sci. ({\em in press}).

\bibitem{kim97}
Y.J. Kim, Y. Gao, and S.A. Chambers, Appl. Surf. Sci. {\bf 120}, 250 (1997);
and references therein.

\bibitem{ashcroft81}
N.W. Ashcroft and N.D Mermin,
{\em Solid State Physics}, CBS Publishing, Tokio (1981).

\end{references}
\end{document}